\documentclass[12pt,draftclsnofoot,journal,onecolumn]{IEEEtran}

\usepackage{graphicx}
\usepackage{booktabs}
\usepackage{subfigure}
\usepackage{color}
\usepackage{multicol}
\usepackage{amssymb}
\usepackage{multirow}
\usepackage{amsmath}
\usepackage{bm}
\usepackage{cite}
\usepackage{gensymb}
\usepackage{amsfonts}
\usepackage{mathrsfs}
\usepackage{amsmath}
\usepackage{algorithm}
\usepackage{algorithmic}
\usepackage{amsthm}

\usepackage{tabularx}
\usepackage{balance}
\usepackage{mathrsfs}
\usepackage{array}
\usepackage{multirow}
\newcommand{\PreserveBackslash}[1]{\let\temp=\\#1\let\\=\temp}
\newcolumntype{C}[1]{>{\PreserveBackslash\centering}p{#1}}
\newcolumntype{L}[1]{>{\PreserveBackslash\raggedright}p{#1}}
\newcolumntype{R}[1]{>{\PreserveBackslash\raggedleft}p{#1}}
\usepackage[flushleft]{threeparttable}
\newcommand{\mb}[1]{{  \mathbf  #1}}  

\begin{document}
	
	\bibliographystyle{IEEEtran} 
	
	\title{Transmissive RIS for 6G Communications: Design, Prototyping, and 
	Experimental Demonstrations}
	\author{\IEEEauthorblockN{ Junwen Tang, Mingyao Cui, Shenheng Xu, 
	\emph{Member, IEEE}, Linglong Dai, \emph{Fellow, IEEE}, Fan Yang, 
			\emph{Fellow, IEEE}, and Maokun Li, \emph{Senior Member, IEEE}}

	\thanks{		Part of this work has been accepted by 2020 IEEE 
	International Symposium on Antennas and Propagation and North American 
	Radio Science Meeting \cite{1}.
	
	All the authors are with the Department of Electronic 
	Engineering, Tsinghua University, Beijing 100084, China (e-mails: 
	tjw17@mails.tsinghua.edu.cn; cmy20@mails.tsinghua.edu.cn; 
	\{shxu, daill, fan\_yang, maokunli\}@tsinghua.edu.cn).
}
\vspace{-4mm}
}
	\maketitle
	\IEEEpeerreviewmaketitle
	\begin{abstract}
Reconfigurable intelligent surface (RIS) has been widely considered as a 
key technique to improve spectral efficiency for 6G communications. 
Compared with most existing research that only focuses on the reflective RIS, 
the design and prototyping of a novel transmissive RIS are presented in this 
paper, and its enhancement to the RIS-aided communication system is 
experimentally demonstrated. The 2-bit transmissive RIS element utilizes the 
penetration structure, which combines a 1-bit current reversible dipole and a 
90° digital phase shifter based on a quadrature hybrid coupler. A transmissive 
RIS prototype with 16×16 elements is designed, fabricated, and measured to 
verify the proposed design. The measured phase shift and insertion loss of the 
RIS element validate the 2-bit phase modulation capability. Being illuminated 
by a horn feed, the prototype achieves a maximum broadside gain of 22.0 dBi at 
27 GHz, and the two-dimensional beamforming capability with scan angles up to 
±60° is validated. The experimental results of the RIS-aided communication 
system verify that by introducing the extra gain and beam steering capability, 
the transmissive RIS is able to  achieve a higher data rate, reduce the 
transmit power, improve the transmission capability through obstacles, and 
dynamically adapt to the signal propagation direction.
	\end{abstract}
	
	\begin{IEEEkeywords}
		Transmissive RIS, 2 bit phase shifter, prototyping \vspace{-3mm}
	\end{IEEEkeywords}	

\section{Introduction} \label{sec:1}
With the explosive growth of the emerging new applications, such as holographic 
video and meta-verse,  6G is expected to achieve a 10$\times$ increase in 
spectral efficiency compared to 5G \cite{2}, \cite{3}. To 
realize this goal, the reconfigurable intelligent surface (RIS) has become an 
essential candidate \cite{4}. By adjusting the phase and amplitude 
responses to an 
incident electromagnetic wave, RIS can achieve dynamic beamforming to enhance 
the spectral efficiency and overcome blockage \cite{5}.
 
Nevertheless, existing research contributions mainly consider the employment of 
reflective RIS \cite{6, 7, 8}, which 
may result in 
coverage holes 
in a cell. 
Specifically, for reflective RIS, the base station and user need to be 
located on the same side of the RIS, which brings extra geographical 
constraints on the physical topology \cite{9}. For instance, 
reflective RIS is 
difficult to assist communication between a transmitter outside a vehicle and a 
receiver inside it. To cope with this problem, transmissive RIS 
has been recently proposed \cite{10}. Rather than being reflected, 
signals can 
transmit through a transmissive RIS to form directional beams. In this way, 
transmissive RIS can potentially fill in the coverage holes of reflective RIS.

The phase reconfigurability is the most essential capability for a transmissive 
RIS. Solid-state electronic devices are commonly integrated in each constituent 
RIS element to dynamically control its phase response. Continuous phase shifts 
can be realized using analog-type devices like varactor diodes 
\cite{10, 11, 12, 13, 15, 14, 16, 17}. The 
phase shift range usually exceeds 360° so that the phase errors are negligible. 
However, the transmissive RIS has a very high transmission insertion loss, 
which can be up to 5.7 dB \cite{13}. As a consequence, the 
aperture 
efficiency of the RIS is considerably reduced. At high frequencies, switch-type 
devices are widely used to produce discrete phase shifts, including PIN 
diodes, RF MEMS switches, and 
etc   
\cite{13,14,15,16,17}. Most existing designs focus on 1-bit phase 
reconfigurability, wherethe 
design and fabrication difficulties are manageable 
\cite{18,20,21,23,24,22,25,26,19}. For 
example, a 
classic design of an O-slot patch loaded with two PIN diodes  was developed in 
\cite{18} to achieve a stable 180° phase shift in a large bandwidth using 
the 
current reversal mechanism. The minimum measured insertion loss is 1.7/1.9 dB 
at 9.8 GHz. Several other attempts were made to obtain a 2-bit phase resolution 
\cite{27,28}. Specifically, a monolithically fabricated 2-bit prototype using 
MEMS 
switches \cite{27} was able to produce 4 phase states, but the measured 
insertion 
loss is 4.2–9.2 dB at Ka band. A compact 2-bit element using a modified O-slot 
patch \cite{28} was designed, fabricated and measured. The measured insertion 
loss 
is 1.5–2.3 dB at 29.0 GHz, and the measured 3-dB transmission bandwidth is 
10.1–12.1\% for four phase states.

The phase quantization errors associated with discrete phase shifts inevitably 
introduce performance degradation \cite{29,30,31}. It is observed that the 
1-bit 
element designs suffer from 3–4 dB insertion loss attributed to the coarse 
phase resolution, which results in a low aperture efficiency. The sidelobe 
levels are significantly higher, sometimes even causing unwanted grating lobes. 
By contrast, 2-bit designs are considered a well-balanced choice between the 
element performance and the design complexity. The phase quantization loss can 
be greatly reduced to less than 1 dB, and the sidelobe envelop can be 
significantly improved as well.

Nonetheless, there are only a few 2-bit designs in the literature. One of the 
important reasons is that, compared to 1-bit designs, more transition 
structures, electronic devices and DC/RF biasing circuits are needed to achieve 
four stable phase states. The conventional design approach solely relies on the 
two-dimensional unit-cell footprint because of the printed-circuit board (PCB) 
fabrication process. It becomes extremely challenging to lay out a variety of 
constituent element parts in a greatly limited sub-wavelength array grid. 
Consequently, more numbers of stacked layers are used, which results in the 
higher fabrication complexity and cost. To the best of our knowledge, the 
design, fabrication, and measurement of the 2-bit transmissive RIS have not 
been well studied.

In this paper, the 2.5-D penetration structure \cite{32} is exploited to make 
the 
best of the longitudinal space in the propagation direction to develop the 
2-bit transmissive RIS. Specifically, our contributions are summarized 
below. 

\begin{itemize}
	\item Firstly, a 90° digital phase shifter with twoPIN diodes based on a 
	quadrature hybrid coupler \cite{33} is proposed and realized.  Combined 
	with a 
	1-bit current reversible dipole, 2-bit phase resolution is achieved at the 
	millimeter wave (mmWave) band. In addition to some preliminary results 
	reported in the authors’ conference paper \cite{1}, the 2-bit RIS element 
	is 
	further optimized and more detailed analysis is presented in this journal 
	paper. 
	\item Furthermore, a prototype with 16×16 RIS elements and a logic circuit 
	control board is fabricated and measured. The measured results show that 
	the prototype can achieve a maximum broadside gain of 22.0 dBi at 27 GHz, 
	with the corresponding aperture efficiency of 25.3\%. Moreover, the 
	measured radiation performances for scan angles are presented as well, 
	demonstrating the beamsteering capability in a large angle up to 60°. 
	\item Finally, a transmissive RIS-aided mmWave wireless communication 
	prototype is developed to demonstrate the performance of the fabricated 
	2-bit transmissive RIS. The experimental results verify that the 2-bit 
	transmissive RIS can achieve a higher data rate, reduce the transmit power, 
	improve the transmission capability through obstacles, and dynamically 
	adapt to the signal propagation direction.
\end{itemize}

\emph{Organization}: The reminder of this paper is organized as follows. The 
communication model of transmissive RIS is introduced in Section II. Then, in 
Section III, we describe the structure, biasing circuits, and simulated 
performance of the element of the designed transmissive RIS. Section IV 
provides the experimental 
results for the 16×16 elements RIS prototype in the microwave anechoic chamber. 
A transmissive RIS-aided wireless communication prototype is set up to further 
measure the performance of the proposed RIS in Section V. Finally in Section 
VI, conclusions are drawn.

\section{System Model} \label{sec:2}
In this section, we describe the system model of the transmissive 
RIS-aided communication system.

\subsection{System Model} \label{sec:2-1}

 	\begin{figure}
	\centering
	{\includegraphics[width=3.5in]{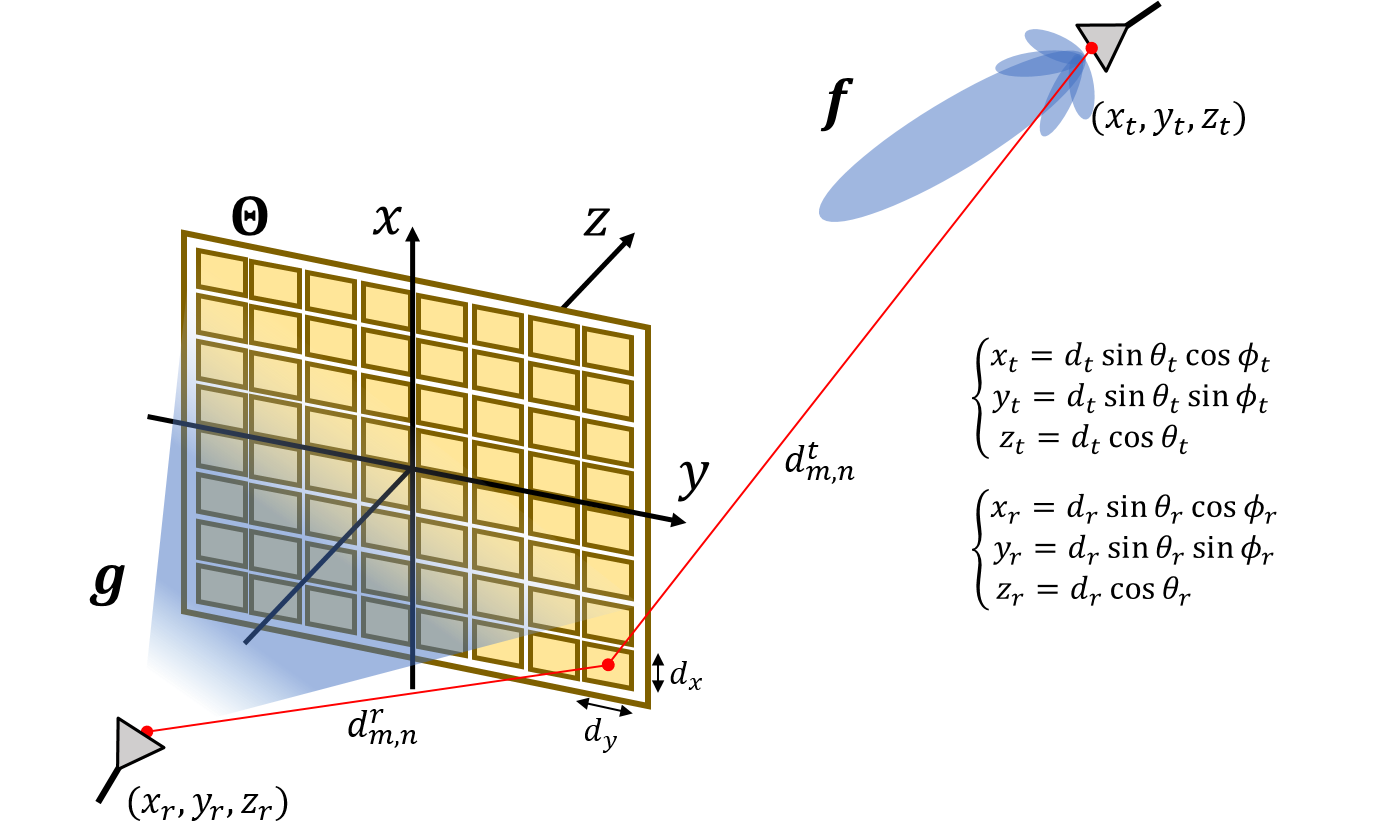}}
	\caption{  System model of transmissive RIS-aided communication systems.
	} 
	\label{fig1}
	\vspace*{-1em}
\end{figure}

We consider that the transmitter (Tx) employs a horn antenna to communicate 
with a single-antenna receiver (Rx), where an N-element transmissive RIS is 
used for enhancing the signal power. As shown in Fig. \ref{fig1}, the RIS is a 
uniformly 
planar array (UPA) placed in the x-y plane, and its geometric center is located 
at the origin of the coordinate system. Let $N_x$ denote the number of antennas 
on the row and $N_y$ denote the number 
of antennas on the column, with $N = N_xN_y$. The antenna spacing 
along the $x$ axis and $y$ axis are  $d_x$ and $d_y$  respectively, which are 
usually half of the wavelength. 
The coordinates of the Tx and Rx are  $(x_t,y_t,z_t)$  and $(x_r, y_r, z_r)$. 
The coordinate of the $(m, n)$-th antenna in the $m$-th row and  $n$-th column 
of the RIS is $(\delta_md_x, \delta_nd_y, 0)$, where  $\delta_m = m - (N_x - 
1)/2$ and $\delta_n = n - (N_y - 1)/2$  with  $m \in\{0,1,\cdots,N_x - 1\}$ and 
$n \in \{0,1,\cdots,N_y - 1\}$. We use the symbols $d_t$, $\theta_t$, and 
$\phi_t$ to denote the distance, the azimuth angle, and the elevation angle 
between the center of the RIS and the Tx. The relationship between $(x_t, y_t, 
z_t)$  and $(d_t, \theta_t, \phi_t)$  is shown in Fig. \ref{fig1}. Similarly, 
we use the 
symbols $d_r$, $\theta_r$, and $\phi_r$ to denote the distance, the azimuth 
angle, and the elevation angle between the center of the RIS and the Rx, where 
the relationship between   $(x_r, y_r, z_r)$  and $(d_r, \theta_r, \phi_r)$  is 
shown in Fig. \ref{fig1} as well.

We denote $\mb{f} \in \mathbb{C}^{N\times 1}$  as the Tx-RIS channel, 
and denote $\mb{g} \in \mathbb{C}^{N\times 1}$ as the RIS-Rx channel. 
Since this paper aims to evaluate the function of the transmissive RIS, the 
direct link between the Tx and Rx is ignored in this paper. Denote 
$[\mb{g}]_{m,n}$   and $[\mb{f}]_{m,n}$ as 
the channel from the $(m, n)$-th antenna of the RIS to the Rx and Tx, 
respectively. Then, the free space channel model  is adopted to represent   
$[\mb{g}]_{m,n}$   and $[\mb{f}]_{m,n}$  as \cite{4}
\begin{align}
[\mb{g}]_{m,n} = \sqrt{\frac{\lambda G_g F_g(\theta_r, \phi_r)}{4\pi}} 
\frac{e^{-j\frac{2\pi}{\lambda}d_{m,n}^r}}{d_{m,n}^r}, \\
[\mb{f}]_{m,n} = \sqrt{\frac{\lambda G_f F_f(\theta_t, \phi_t)}{4\pi}} 
\frac{e^{-j\frac{2\pi}{\lambda}d_{m,n}^t}}{d_{m,n}^t}, 
\end{align}
where  $\lambda$ represents the wavelength, $d_{m,n}^r$  and $d_{m,n}^t$
denote the distance from the $(m, n)$-th antenna of the RIS to the Rx and Tx, 
respectively. Moreover,  $G_g$ and $F_g(\theta_r, \phi_r)$
denote the antenna gain and the normalized power radiation pattern of the side 
of RIS facing the receiver, while  $G_f$  and $F_f(\theta_t, \phi_t)$  
correspond to those of the side of RIS facing the transmitter \cite{4}.

Let $s$ denote the transmitted symbol with $\|s\|_2 = 1$. 
We use symbols $G_t$, $F_t$, $P_t$, $G_r$, $F_r$ to represent the transmitter 
antenna gain, the transmit power, the normalized radiation power of the 
transmitter, 
the receiver antenna gain, and the normalized radiation power of the receiver. 
Then 
the noiseless received signal by the RX can be modeled as \cite{4}
\begin{align}\label{eq:channel}
y &= \sqrt{P_tGF}\sum_{m=0}^{N_x - 1}\sum_{n=0}^{N_y - 
	1}[\mb{g}]_{m,n}
[\mb{f}]_{m,n}\Gamma_{m,n}e^{j\phi_{m,n}}s, \notag\\
&=\frac{\sqrt{P_tGF\lambda}}{4\pi}\sum_{m=0}^{N_x - 1}\sum_{n=0}^{N_y - 
1}\frac{\Gamma_{m,n}}{d_{m,n}^td_{m,n}^r}e^{j\left(\phi_{m,n}
 - \frac{
d_{m,n}^t+d_{m,n}^r}{\lambda}\right)}s,
\end{align}
where $\Gamma_{m,n}\in[0,1]$ is transmission loss of the $(m,n)$-th antenna of 
the RIS, $\phi_{m,n}\in[0, 2\pi]$ denotes the phase shift of the $(m,n)$-th 
antenna of the RIS, $G = G_tG_fG_gG_r$, and $F = F_tF_f(\theta_r, 
\phi_r)F_g(\theta_r, \phi_r)F_r$.

\subsection{Beamforming for the Transmissive RIS} \label{sec:2-2}

For the desired receiver, the transmissive RIS is designed for maximizing the 
received signal energy through beamforming. The noiseless 
received signal energy can be presented as 
\begin{align}
P_r = \frac{P_tGF\lambda^2}{16\pi^2}\left|\sum_{m=0}^{N_x - 1}\sum_{n=0}^{N_y - 
	1}
\frac{\Gamma_{m,n}}{d_{m,n}^td_{m,n}^r}e^{j\left(\phi_{m,n}
	- \frac{
		d_{m,n}^t+d_{m,n}^r}{\lambda}\right)}\right|^2.
\end{align}
To maximize $P_r$,  it is obvious that the optimal $\phi_{m,n}$ is 
\begin{align}
\bar{\phi}_{m,n} = \mod\left(C + \frac{
	d_{m,n}^t+d_{m,n}^r}{\lambda}, 2\pi\right).
\end{align}
Here, $C$ is an arbitrary constant.  
From (5), the phase shift $\phi_{m,n}$  can be obtained according to 
$d_{m,n}^t$  and $d_{m,n}^r$. In our practical employment,  the RIS is deployed 
nearby the Rx and far away from the Tx. That is to say, the Rx is within the 
near-field range of the RIS, while the Tx is within the far-field range of the 
RIS. Based on this assumption, the distance $d_{m,n}^r$ can be derived 
according to the 
spherical wave model as 
\begin{align}
d_{m,n}^r = \sqrt{(x_r - \delta_md_x)^2 + (y_r - \delta_nd_y)^2 + z_r^2},
\end{align}
while the distance $d_{m,n}^t$  can be simplified by the planar wave model as
\begin{align}
d_{m,n}^t &= \sqrt{(x_t - \delta_md_x)^2 + (y_t - \delta_nd_y)^2 + z_t^2} 
\notag\\
&\approx r_t - \delta_md_x\sin\theta_t\cos\phi_t - \delta_nd_y\sin\theta_t
\sin\phi_t.
\end{align}
Therefore, the optimal solution to $\phi_{m,n}$ can be obtained accordingly. 
Then, for practical implementation of RIS, the limited resolution of the phase 
shift has to be considered. If there are $b$ bits for quantized phase shifter, 
then the feasible set of phase shift is 
\begin{align}
\mb{\Phi}_b = \left\{0, \frac{1}{2^{b-1}}\pi,\cdots, 
\frac{2^{b}-1}{2^{b-1}}\pi\right\}.
\end{align}
Finally, the practical phased shift on a RIS element is given by
\begin{align}
\tilde{\phi}_{m,n} = \arg\min_{\phi \in \mb{\Phi_b}} |\phi-\bar{\phi}_{m,n}|.
\end{align}
In our system, we design a transmissive RIS with $b = 2$ bits. Notice that the 
transmission model and the phase shift design mathematically have 
no difference with the reflective RIS-aided communication system.  
Nevertheless, the application scenarios and implementation and of them are 
quite different. For instance, a transmissive RIS can be deployed on a vehicle 
window to enhance communication between inside and outside the vehicle. In the 
following sections, we will elaborate on the proposed transmissive RIS design 
in our communication system.

\section{Proposed RIS Element Design}
We describe the proposed transmissive RIS element in this section, 
including element design, 90° phase shifter exploited in the element, biasing 
circuits, and element performance in simulation.
\subsection{Element Design}
 	\begin{figure}
	\centering
	{\includegraphics[width=3.5in]{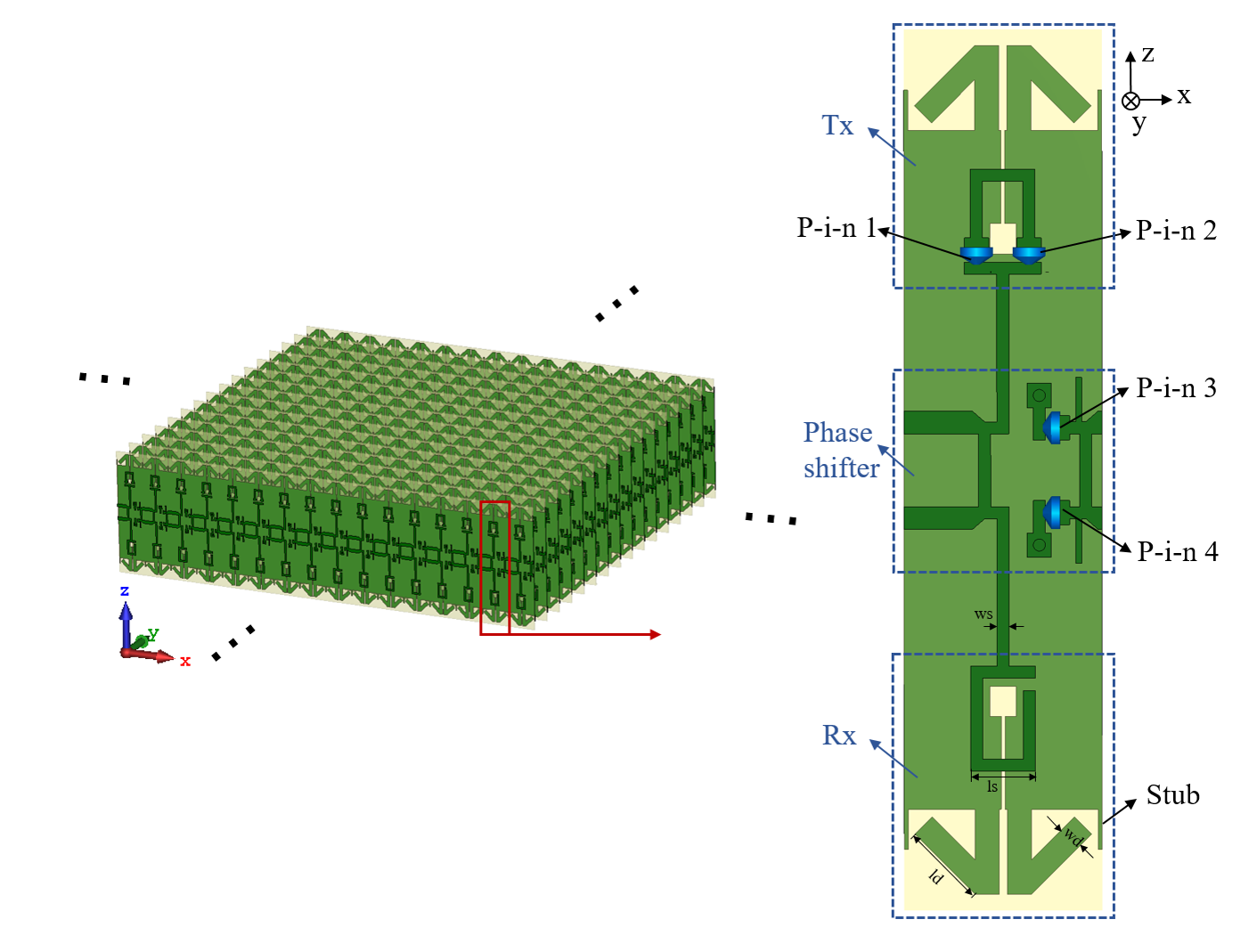}}
	\caption{ Proposed 2.5-D structure of the transmissive RIS for mmWave 
	communications.} 
	\label{fig2}
	\vspace*{-1em}
\end{figure}
{
\begin{table}[]
	\caption{Simulated Performance of the 2-bit RIS Element}
	\centering
	\begin{tabular}{lccc}
		\toprule
		\multicolumn{1}{c}{State} & \begin{tabular}[c]{@{}c@{}}Element loss\\ 
		@26.5 GHz\end{tabular} & 
		\multicolumn{1}{c}{\begin{tabular}[c]{@{}c@{}}Phase shift\\ @26.5 
		GHz\end{tabular}} & \multicolumn{1}{c}{\begin{tabular}[c]{@{}c@{}}3-dB 
		trans.\\ bandwidth\end{tabular}} \\
		\midrule
		1 (state-0$^{\circ}$)               & 1.1 
		dB                                                            & 
		-141.2$^{\circ}$                                                        
		         
		             &
		 23.6-28.9 GHz 
		(20.1\%)                                                              \\
		\midrule
		2 (state-90$^{\circ}$)              & 1.3 
		dB                                                            & 
		-56.8$^{\circ}$                                                         
		          
		             &
		 24.8-28.6 GHz 
		(14.2\%)                                                              \\
		\midrule
		3 (state-180$^{\circ}$)             & 1.1 
		dB                                                            & 
		34.9$^{\circ}$                                                          
		          
		             &
		 23.5-28.6 GHz 
		(19.6\%)                                                              \\
		\midrule
		4 (state-270$^{\circ}$)             & 1.5 
		dB                                                            & 
		129.0$^{\circ}$                                                         
		          
		             &
		 24.6-29.0 GHz 
		(16.4\%)  \\ 
		\bottomrule                                                         
	\end{tabular}
\end{table}
}
Fig. \ref{fig2} illustrates the proposed 2.5-D structure of the 2-bit 
transmissive RIS. Each RIS element consists of a number of one-dimensional 
subarrays aligned in parallel. The constituent element utilizes the 
penetration structure, which 
is composed of a RIS-side Rx and a RIS-side Tx  to receive and transmit the RF 
signals impinging on the RIS, respectively, and a microstrip transmission line 
to convey the signal through the RIS aperture. Instead of laying them out in 
the aperture plane, two vertical dipoles \cite{35} are employed and arranged 
along 
the propagation direction to circumvent the space constraint imposed by the 
unit cell footprint. The slightly extended longitudinal space easily 
accommodates a 90° digital phase shifter between the RIS-side Rx and the 
RIS-side Tx. Combined with the 1-bit current reversible dipole, the 2-bit 
transmissive RIS element can be readily realized.

The 2.5-D structure slightly increases the aperture profile by 1~2 wavelengths, 
but it is still much smaller compared with the aperture size. Especially at 
mmWave frequencies, the aperture profile is merely a couple of centimeters and 
will not considerably undermine the low profile and conformal features of the 
RIS. Each subarray can be easily fabricated on a single layer of substrate, 
thus avoiding the complicated multilayer PCB process and lowering the overall 
cost. It is also a remarkable fact that the $//$-type 2.5-D structure can be 
further developed into a $\#$-type or a $\Delta$-type formation where two or 
three sets 
of subarrays can be arranged in the same aperture, so that the 2.5-D unit cell 
space can be more fully exploited for more constituent element parts.  Hence, 
more advanced features such as dual-polar or dual-band designs can be 
accomplished.

The linearly incident wave propagating along the z axis is received by the 
passive RIS-side Rx composed of a vertical dipole. Then, the received signal is 
converted into a guided wave and conveyed along the microstrip transmission 
line. The phase shifter structure based on the quadrature hybrid coupler is 
employed. Instead of the varactor diodes used in the analog phase shifter 
design [33], it is terminated with reflection-type loads composed of two p-i-n 
diodes (model MADP-000907-14020) to provide a 0°/90° digital phase shift. The 
two loaded p-i-n diodes are always tuned at the same ON- or OFF-states, 
resulting in a 90° phase shift. The RIS-side Tx is similar to the RIS-side Rx, 
but the vertical dipole structure \cite{35} is modified to integrate two p-i-n 
diodes (MADP-000907-14020) that are symmetrically placed in the feeding 
microstrip line. They are alternatively turned ON or OFF so that the excitation 
current flows in opposite directions. Consequently, the reversible current flow 
produces a stable 180° phase shift without significantly affecting magnitude 
response. Finally, the RIS-side Tx re-radiates the guided wave into the free 
space with four stable phase states of 0°/90°/180°/270°, and hence, the 2-bit 
phase resolution is obtained.

The frequency of the designed transmissive RIS element is 26.5 GHz, and the 
unit 
cell size is $4.9\times4.9$ $\text{mm}^2$. The proposed 2-bit element is 
printed 
on a single layer of Taconic TLX-8 substrate ($\epsilon_r$ = 2.55, 
$\tan\delta$= 0.0019, $h$ = 
0.254 mm). The microstrip line and the 90° digital phase shifter are on the top 
of the substrate, while the vertical dipoles and the ground plane are arranged 
on the bottom of the substrate. The total profile of the element is 21.8 mm, 
corresponding to 1.9$\lambda$ at the design frequency. The following 
geometrical 
parameters are optimized after a comprehensive parametric study: $l_d$ = 2.1 
mm, 
$wd$ = 0.6 mm, $l_s$ = 1.6 mm, $w_s$ = 0.3 mm.

\subsection{90° Digital Phase Shifter}
 	\begin{figure}
	\centering
	{\includegraphics[width=5in]{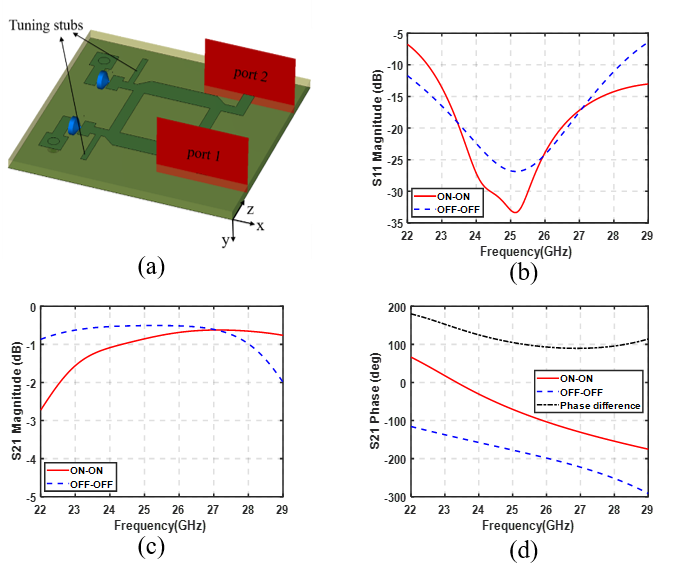}}
	\caption{ Proposed 90° digital phase shifter based on a quadrature hybrid 
	coupler: (a) the illustrative structure; (b) simulated reflection 
	magnitude, (c) transmission magnitude, and (d) transmission phase.} 
	\label{fig3}
	\vspace*{-1em}
\end{figure}
The proposed 90° digital phase shifter integrated in the 2-bit transmissive RIS 
element is designed to provide 0°/90° phase switching, as shown in Fig. 
\ref{fig3} (a). 
It is a reflection-type phase shifter based on the quadrature coupler 
terminated with two reflective loads, which are composed of p-i-n diodes and 
short-circuited stubs. The realization of 90° phase shift is by tuning the two 
p-i-n diodes to work at ON-ON or OFF-OFF state, respectively. Two tuning stubs 
are introduced to maintain a stable desired phase shift by tuning their 
lengths. The digital phase shifter is modeled and simulated with a periodic 
boundary condition and a wave-port excitation. As can be observed from the 
simulation results plotted in Fig. \ref{fig3} (b) , a 10-dB return loss 
bandwidth of 
22\% (22.6-28.2 GHz) is achieved. 
Good impedance matching performance minimizes 
the insertion loss to less than 1 dB within the frequency band from 24.2 to 
28.0 GHz. Meanwhile, from the curves plotted in Fig. \ref{fig3} (d) , we can 
find that 
the phase difference is close to 90° in a wide frequency band. We can conclude 
that the phase shifter is capable of generating desired 90° phase shift by 
electrically adjusting the two p-i-n diodes, thus achieving the 0°/90° digital 
phase switching capability with the low insertion loss.

\subsection{Performance of the designed Transmissive RIS Element}
 	\begin{figure}
	\centering
	\subfigure[]
	{\includegraphics[width=3in]{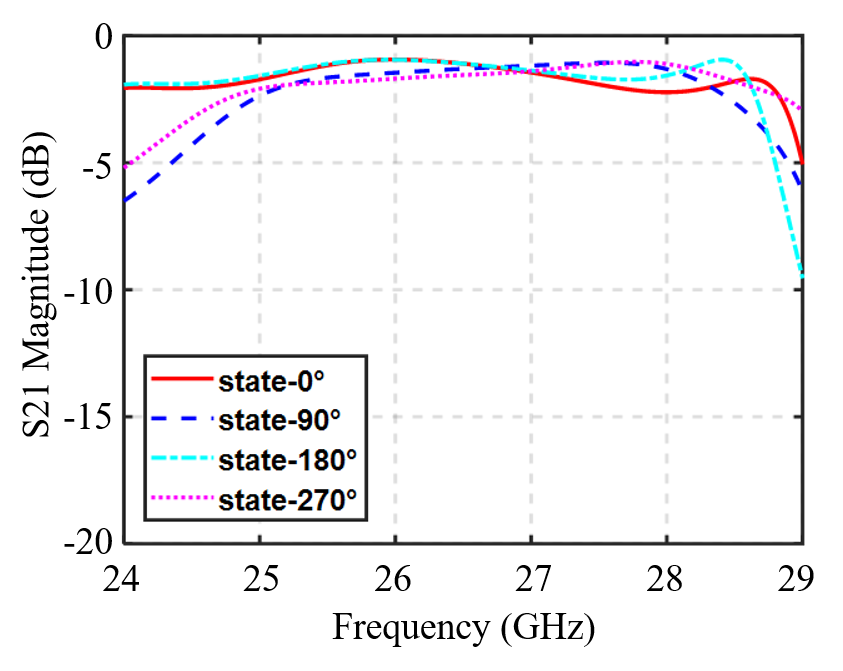}}
	\subfigure[]
	{\includegraphics[width=3in]{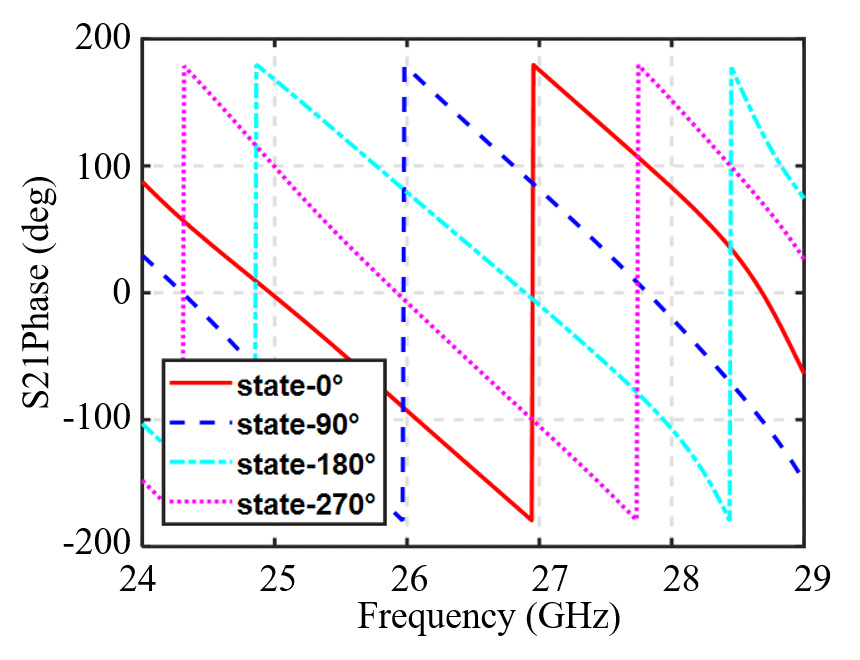}}
	\caption{ Simulated transmission (a) magnitude and (b) phase of the 
	proposed 2-bit RIS element.} 
	\label{fig4}
	\vspace*{-1em}
\end{figure}
We simulate the proposed 2-bit RIS element with a periodic boundary condition 
to mimic an infinite array environment and the mutual 
coupling effect is also considered. The element is simulated. The simulated 
element performances 
under normal incidence for the four working states are plotted in Fig. 
\ref{fig4}. The 
transmission magnitude responses at 26.5 GHz are summarized in Table I. It can 
be seen that the average transmission insertion loss of the four states is 
around 1.3 dB. Moreover, the insertion losses are maintained below 3 dB ranging 
from 24.8 to 28.6 GHz, corresponding to the 3-dB transmission bandwidth of 
14.2\%. The phase curves of the four states are roughly parallel with around 
90° phase difference in the operating frequency band. Hence, by switching the 
p-i-n diodes alternatively, the RIS element successfully achieves 2-bit phase 
tuning capability.

\section{Ris Design, Fabrication, and Measurement}
In this section, a transmissive RIS prototype composed of $16\times16$ proposed 
elements is designed, fabricated, and measured to experimentally validate the 
performance of the proposed RIS element.

 	\begin{figure}
	\centering
	{\includegraphics[width=5in]{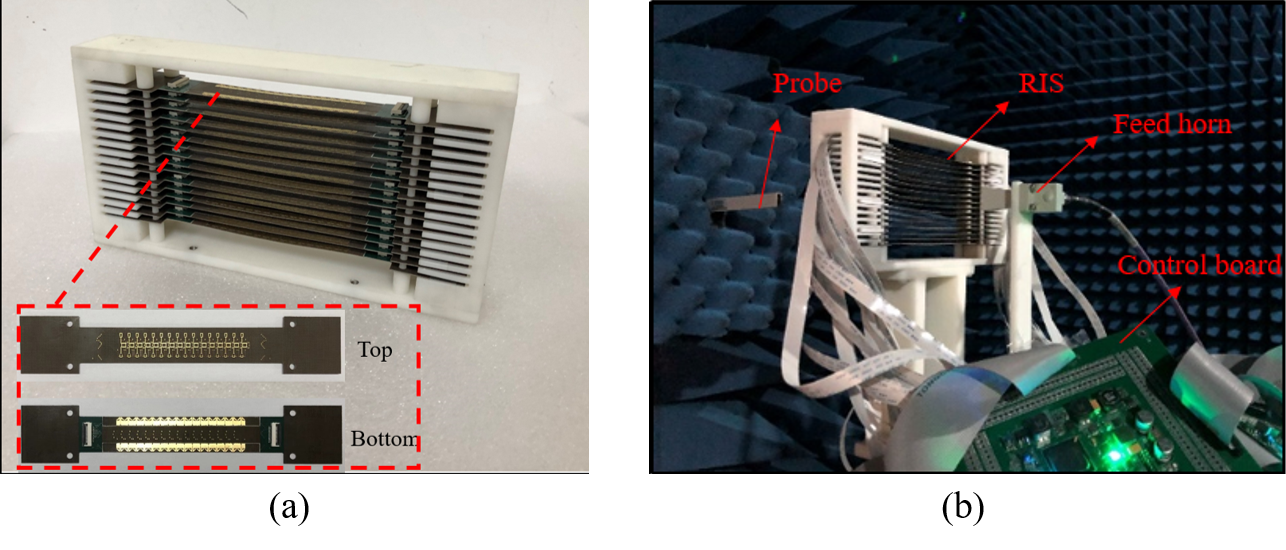}}
	\caption{ Photography of (a) the fabricated transmissive RIS prototype, (b) 
	the measurement setup in the microwave anechoic chamber.} 
	\label{fig5}
	\vspace*{-1em}
\end{figure}

The photographs of the fabricated RIS prototype and the measurement setup in 
the microwave anechoic chamber are presented in Fig. \ref{fig5}. Taking 
advantage of 
the  PCB technology, the RIS is easy to fabricate. The RIS is space-fed by a 
linearly polarized horn. The effective aperture size of the RIS prototype is 
$78.4\times78.4$ $\text{mm}^2$. The feeding horn illuminates the array at an 
optimized distance of 50 mm. To achieve the goal of phase reconfigurability, 
each element of the array needs to be controlled independently. There are 512 
biasing lines since each element has two biasing lines to apply DC voltages to 
four p-i-n diodes. Two logic circuit controlling boards are attached to supply 
independent DC voltages of 512 channels, and they are connected to the biasing 
lines through several connectors.

 	\begin{figure}
	\centering
	\subfigure[]
	{\includegraphics[width=3in]{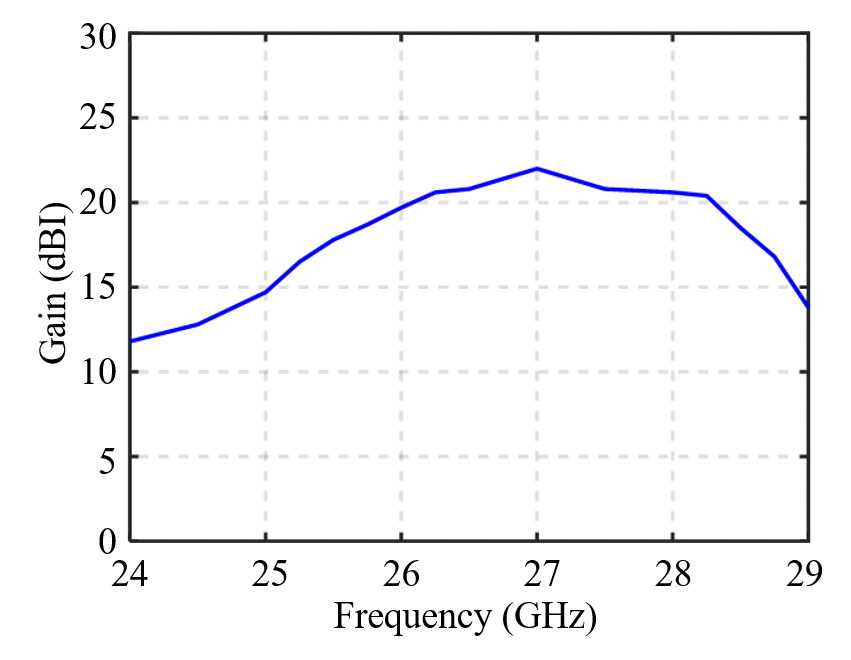}}
	\subfigure[]
	{\includegraphics[width=3in]{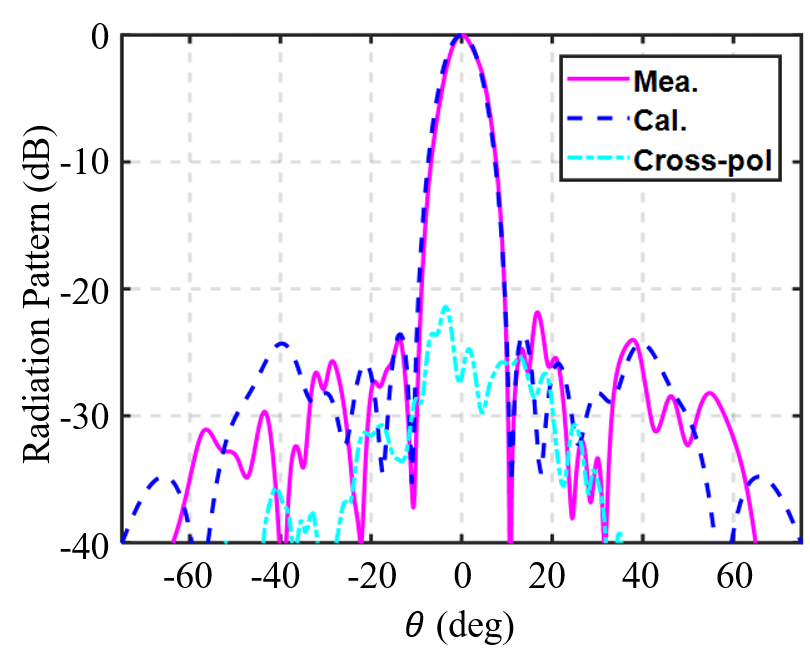}}
	\caption{Performance of the broadside beam: (a) measured gain bandwidth, 
	(b) calculated and measured normalized radiation patterns at 27.0 GHz.} 
	\label{fig6}
\end{figure}
We measure the fabricated prototype in a planar near-field anechoic chamber. 
From the measured performance of the broadside beam depicted in Fig. 
\ref{fig6}, a peak 
gain of 22.0 dBi is obtained at 27.0 GHz, which is 9.3 dB higher than that of 
the feeding horn. The corresponding aperture efficiency is 25.3\%. The slight 
frequency shift to higher frequencies is mainly attributed to the tolerance in 
fabrication and the supporting structure. The measured radiation pattern is in 
good agreement with the theoretical prediction based on the array theory. The 
measured sidelobe level is below -22 dB, remarkably lower than that of the 
1-bit designs, which is mainly attributed to the finer phase resolution of the 
2-bit RIS. The measured cross-polarization level is -21.3 dB. 

 \begin{figure}
	\centering
	{\includegraphics[width=5in]{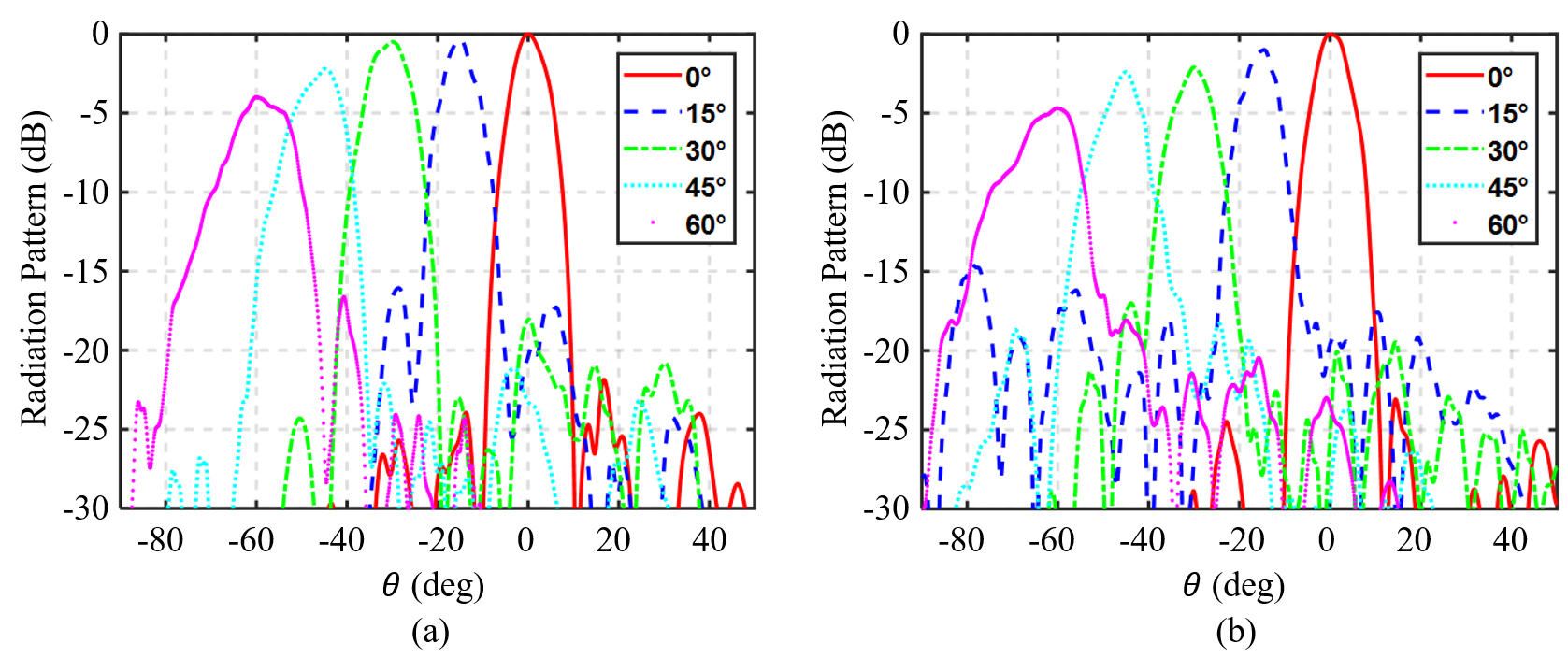}}
	\caption{Measured radiation patterns of scanned beams at 27.0 GHz in (a) 
	E-plane, (b) H-plane.} 
	\label{fig7}
\end{figure}

The radiation behaviors of the scanned beams are measured as well to 
experimentally verify the dynamic beam scanning capability of the transmissive 
RIS in two-dimensional space. By controlling the phase of each RIS element, the 
transmitted focused beam can be formed by the RIS in the given direction. As 
shown in Fig. \ref{fig7}, the measured scanned beams in the angular region from 
0° to 
-60° show accurate beam pointing and well-defined patterns in both E-plane and 
H-plane. Note that the beams in the opposite angular region are omitted due to 
the symmetry. The measured 60° scan gain losses are 4.0 dB and 5.0 dB in two 
principal planes, respectively. Table II compares the performance of the 
proposed 2-bit RIS with other RIS designs published in literature. It is 
evident that this work achieves the highest aperture efficiency. 
\begin{table}[]
	\centering
	\caption{Performance Comparison with Other transmissive RIS Designs}
	{
		\begin{tabular}{cccccc}
			\toprule
			\textbf{Ref.}     & \begin{tabular}[c]{@{}c@{}}\textbf{Freq.}\\ 
				\textbf{{[}GHz{]}}\end{tabular} & 
			\begin{tabular}[c]{@{}c@{}}\textbf{Tunable}\\ 
				\textbf{device}\end{tabular} & 
			\begin{tabular}[c]{@{}c@{}}\textbf{Phase}\\ 
				\textbf{resolution}\end{tabular} & 
			\begin{tabular}[c]{@{}c@{}}\textbf{Gain}\\ 
				\textbf{{[}dBi{]}}\end{tabular} & 
			\begin{tabular}[c]{@{}c@{}}\textbf{Aper. eff.}\\ 
				\textbf{{[}\%{]}}\end{tabular} \\
			\midrule
			{[}18{]} & 
			9.8                                                       & 
			p-i-n                                                    & 
			1-bit                                                      & 
			22.7                                                     & 
			15.4                                                                
			    
			\\
			\midrule
			{[}26{]} & 
			34.8                                                      & 
			MEMS                                                     & 
			2-bit                                                      & 
			9.2                                                      & 
			6.2                                                                 
			    
			\\
			\midrule
			{[}27{]} & 
			29                                                        & 
			p-i-n                                                    & 
			2-bit                                                      & 
			19.8                                                     & 
			15.9                                                                
			    
			\\
			\midrule
			Our work & 
			27                                                        & 
			p-i-n                                                    & 
			2-bit                                                      & 
			22.0                                                     & 
			25.3                                                                
			    
			\\\bottomrule   
	\end{tabular}}
\end{table}
\section{Transmissive RIS-Aided Wireless Communication Prototype}
In this section, a transmissive RIS-aided wireless communication prototype is 
set up to further measure the performance of the designed transmissive RIS in 
practical communication scenarios. 

\subsection{Measurement Setup}
 	\begin{figure}
	\centering
	{\includegraphics[width=3.5in]{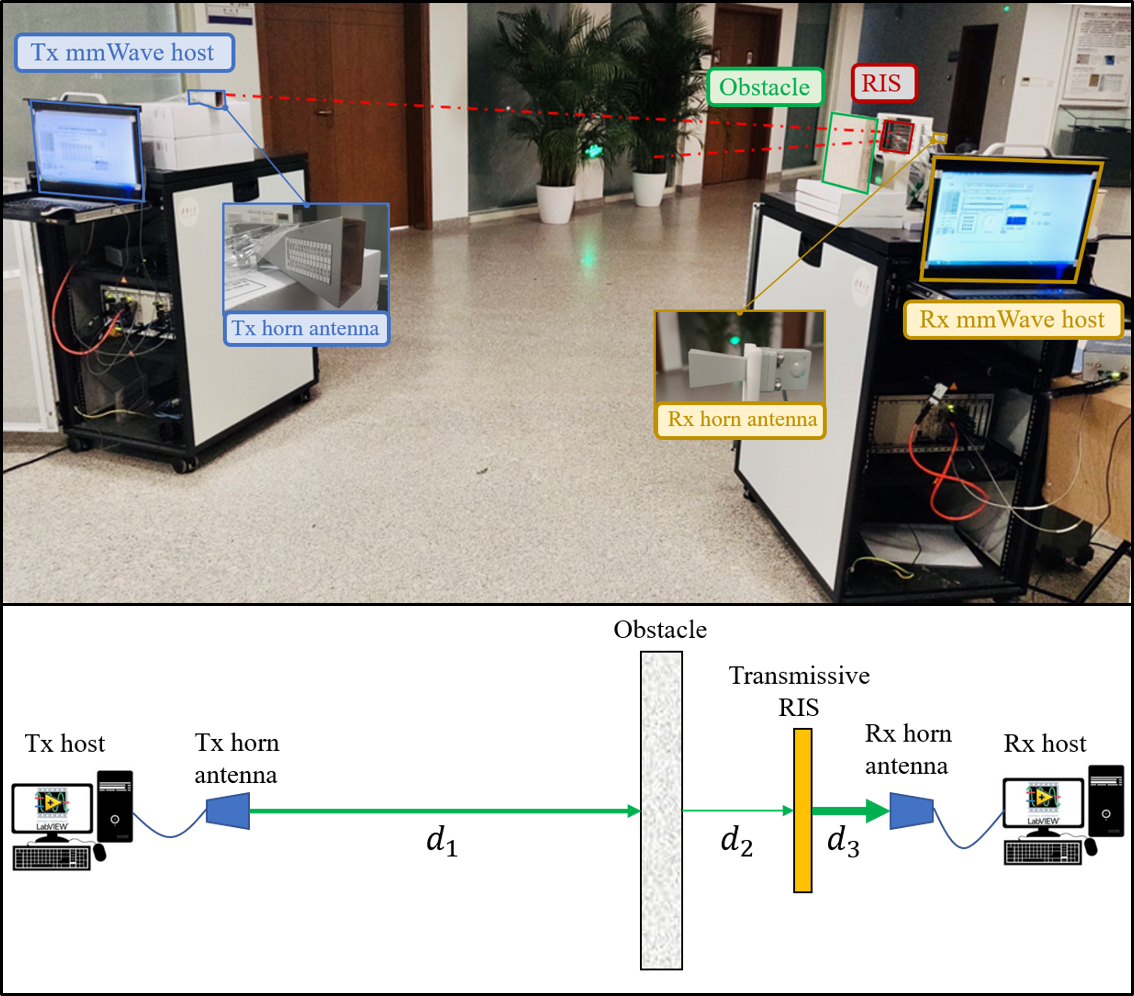}}
	\caption{Transmissive RIS-aided wireless communication prototype.} 
	\label{fig8}
\end{figure}

As illustrated in Fig. \ref{fig8}, we carry out the experiment in an office 
environment. The prototype is composed of the transmitter side, the RIS, and 
the 
receiver side. Both the receiver and the transmitter utilize the 
software-defined 
radio 
(SDR) platform, composed of the PXI hardware architecture and the LabVIEW 
system design software, to realize transmission at the mmWave band. 

\begin{table}[]
		\caption{System Parameters}
		\renewcommand\arraystretch{1.25}
		\centering
	\begin{tabular}{|c|c|cc|c|}\hline
		\specialrule{0em}{1.2pt}{0pt}
		\hline
		\textbf{Parameters}                                                    
		      & \textbf{Value}                   & 
		\multicolumn{2}{c|}{\textbf{Parameters}}

		           &
		 \textbf{Value}          \\ \hline
		\begin{tabular}[c]{@{}c@{}}Carrier\\ 
		frequency\end{tabular}                   & 27.0 GHz                & 
		\multicolumn{1}{c|}{\multirow{2}{*}{\begin{tabular}[c]{@{}c@{}}Antenna 
		\\ gain\end{tabular}}} & 
		Tx                                                         & 
		22.7           \\ \cline{1-2} \cline{4-5} 
		System 
		bandwidth                                                              &
		 800 MHz                 & 
		\multicolumn{1}{c|}{}                                                   
		                      &
		 Rx                                                         & 
		9.2            \\ \hline
		Waveform                                                                
		      & CP-OFDM                 & 
		\multicolumn{1}{c|}{\multirow{4}{*}{\begin{tabular}[c]{@{}c@{}}Trans.\\ 
		\\ RIS\end{tabular}}} & Array 
		gain                                                 & 19.8           
		\\ \cline{1-2} \cline{4-5} 
		FFT 
		size                                                                    
		  &
		 2048                    & 
		\multicolumn{1}{c|}{}                                                   
		                      &
		 Frequency                                                  & 
		22.0           \\ \cline{1-2} \cline{4-5} 
		\begin{tabular}[c]{@{}c@{}}Carrier\\ 
		spacing\end{tabular}                     & 75 KHz                  & 
		\multicolumn{1}{c|}{}                                                   
		                      & \begin{tabular}[c]{@{}c@{}}Phase\\ 
		resolution\end{tabular} & 2 bit          \\ \cline{1-2} \cline{4-5} 
		\multirow{2}{*}{\begin{tabular}[c]{@{}c@{}}ADC/DAC\\ 
		resolution\end{tabular}} & \multirow{2}{*}{14 bit} & 
		\multicolumn{1}{c|}{}                                                   
		                      & Array 
		size                                                 & 16 $\times$ 16 
		\\ \cline{3-5} 
		&                         & \multicolumn{2}{c|}{Tx/Rx 
		height}                                                                 
		                                                         &
		 1.5 m          \\ \hline
		Modulation                                                              
		      & 16-QAM                  & 
		\multicolumn{2}{c|}{$d_1$}                                              
		                                                                        
		           &
		 2.4 m          \\ \hline
		Coding                                                                  
		      & Turbo                   & 
		\multicolumn{2}{c|}{$d_2$}                                              
		                                                                        
		           &
		 0.2 m          \\ \hline
		Code 
		rate                                                                    
		 &
		 1/2                     & 
		\multicolumn{2}{c|}{$d_3$}                                              
		                                                                        
		           &
		 0.05 m         \\ \hline		\specialrule{0em}{1.2pt}{0pt}
		 \hline
	\end{tabular}
\end{table}
The transmitter consists of the transmitter host, the FPGA, the ADC module, 
the mmWave upconverter, and a horn antenna. The transmitter host controls the 
system
parameters, including modulations, the transmit power, the central frequency, 
and so on. A high-speed bitstream is first delivered by the host to 
the FPGA module. The FPGA module realizes complex signal processing, including 
channel coding, constellation mapping, OFDM modulation, and etc. After that, 
the signals are sequentially processed by the ADC and the mmWave upconverter to 
form mmWave analog signals suitable for transmission in wireless channels. 

The transmissive RIS has 256 2-bit elements. It receives the signals from the 
transmitter, tunes the phase of received signals, and then directionally 
transmits them towards the receiver.  Moreover, for
validating that a transmissive RIS is able to overcome obstacles, we can also 
place an obstacle, such as a piece of marble, between the RIS and the base 
station. 

Similar to the transmitter, the receiver is composed of the receiver host, the 
FPGA, a DAC module, a mmWave down converter, and a horn antenna. 
At the receiver side, the Rx horn antenna first receives the signal and 
transforms 
them the digital domain by the mmWave down converter and the DAC module. Then, 
the 
FPGA module is responsible for recovering the bitstream by channel estimation, 
OFDM demodulation, decoding, CRC check and etc. Moreover, according to the 
results of the CRC check, the host obtains the block error rate and calculates 
the data rate. Finally, the received constellation and the data rate are 
illustrated on the screen of the host. 

In Table III, the detailed settings of our mmWave platform are provided. We 
consider a downlink communication scenario. The bandwidth is 800 MHz 
and the carrier frequency is 27.0 GHz. The horn antenna of the transmitter, the 
transmissive RIS, and the
horn antenna of the receiver are on the same height of 1.5 m and are all 
horizontally 
polarized. A piece of marble with thickness of 0.03 m is employed to mimic the 
obstacle. Let $d_1$, $d_2$, and $d_3$ denote the the 
distance between the Tx antenna and the obstacle, the distance between the 
obstacle and the RIS, the distance between the RIS and the Rx antenna. They are 
set as $d_1$ = 2.4 m, $d_2$ = 0.2 m, and $d_3$ = 0.05 m, respectively.
The fabricated 
transmissive RIS works at 27.0 GHz with an additional 9.3 dB array gain when 
being illuminated by the Rx horn antenna. The azimuth angle and the elevation 
angle from   the RIS center to the Tx/Rx are all zero, i.e., $\theta_t 
= \theta_r = \phi_t = \phi_r$ =  0° as shown in Fig. \ref{fig1}. We verify the 
performance of this RIS in terms of array gain and data rate.
\subsection{Array Gain Performance Assessment}
\begin{table}[]\centering
	\caption{Array Gain Performance}
	\begin{tabular}{ccc}
		\toprule
		\textbf{With or without RIS} & \multicolumn{1}{l}{\textbf{Transmit 
		power}} & 
		\multicolumn{1}{l}{\textbf{Data rate}} \\
		\midrule
		$\times$            & 13.6 dBm                           & 1024 
		Mbps                     \\
		\midrule
		$\checkmark$        & 5.4 dBm                            & 1121 
		Mbps \\
		\bottomrule                   
	\end{tabular}
\end{table}

To evaluate the array gain, we remove the obstacle and assess the system 
performance with or without the transmissive RIS. The experiment results are 
presented in Table IV. Specifically, without the transmissive RIS, when the 
transmit power reaches 13.6 dBm, the data rate can reach 1024 Mbps. By 
contrast, to achieve a similar transmission rate of 1121 Mbps, only 5.4 dBm 
transmit power is required with the help of transmissive RIS. It is obvious 
that the 
transmit power is reduced by 8.2 dB. As we measured in the anechoic chamber, 
the RIS can provide a 9.3 dB array gain, which is consistent with the transmit 
power reduction measured by our prototype. Therefore, it is obtained that the 
transmissive RIS-aided communication prototype can reduce the transmit 
power by the magnitude close to the array gain provided by the RIS, while 
maintaining the similar communication performance.
\subsection{Data Rate Assessment}
\begin{table}[]\centering
	\caption{Data Rate Performance}
	\begin{tabular}{ccccc}
		\toprule
		\begin{tabular}[c]{@{}c@{}}\textbf{With or} \\ \textbf{without 
		RIS}\end{tabular} & 
		\begin{tabular}[c]{@{}c@{}}\textbf{With or} \\ \textbf{without 
		obstacle}\end{tabular} & 
		\begin{tabular}[c]{@{}c@{}}\textbf{Azimuth}\\ 
		\textbf{angle}\end{tabular} & 
		\begin{tabular}[c]{@{}c@{}}\textbf{Data} \\ \textbf{rate}\end{tabular} 
		\\
		\midrule
		$\times$                                                       & 
		$\times$                                                            & 
		$0^{\circ}$                                                             
		                                & 1024 
		Mbps                                            \\
		\midrule
		$\checkmark$                                                   & 
		$\times$                                                            & 
		$0^{\circ}$                                                             
		                               & 1683 
		Mbps                                            \\
		\midrule
		$\times$                                                       & 
		$\checkmark$                                                        & 
		$0^{\circ}$                                                             
		                                & 0 
		Mbps                                               \\
		\midrule
		$\checkmark$                                                   & 
		$\checkmark$                                                        & 
		$0^{\circ}$                                                             
		                                 & 1683 
		Mbps        \\
		\bottomrule                                   
	\end{tabular}
\end{table}
To further validate that the fabricated transmissive RIS is able to overcome 
obstacles, we also introduce a piece of marble into the system to evaluate the 
data rate performance. Specifically, depending on whether the obstacle and/or 
the RIS exist, there are in total four communication scenarios. The transmit 
power is fixed to 13.6 dBm. The experiment results are tabulated in Table V. 
Two conclusions can be derived accordingly. One is that the fabricated RIS is 
able to improve the transmission data rate. When the obstacle is removed, the 
introduction of the transmissive RIS increases the data rate from 1024 Mbps to 
1683 Mbps. It is worth mentioning that 1683 Mbps is the highest data rate that 
the communication prototype can support with 16-QAM modulation. The other is 
that the transmissive RIS is able to overcome obstacles because of the 
additional array gain it provides. Specifically, when the direct Tx-Rx link is 
blocked by the marble, data transmission is completely interrupted 
without the help of transmissive RIS. By contrast, the introduction of the 
transmissive RIS 
re-establishes the communication link and achieves the highest 1683 Mbps. This 
is because that the array gain provided by the RIS is higher than the 
attenuation of the obstacle. 
\subsection{Beam Steering Assessment}
\begin{table}[]\centering
	\caption{Beam Steering Evaluation}
	\begin{tabular}{ccccc}
		\toprule
				\begin{tabular}[c]{@{}c@{}}\textbf{With or} \\ \textbf{without 
				RIS}\end{tabular} & 
		\begin{tabular}[c]{@{}c@{}}\textbf{With or} \\ \textbf{without 
				obstacle}\end{tabular} & 
		\begin{tabular}[c]{@{}c@{}}\textbf{Azimuth}\\ 
			\textbf{angle}\end{tabular} & 
		\begin{tabular}[c]{@{}c@{}}\textbf{Data} \\ \textbf{rate}\end{tabular} 
		\\
		\midrule
		$\times$                                                       & 
		$\times$                                                            & 
		$30^{\circ}$                                                            
		                                 & 450 
		Mbps                                            \\
		\midrule
		$\checkmark$                                                   & 
		$\times$                                                            & 
		$30^{\circ}$                                                            
		                                 & 1683 
		Mbps                                            \\
		\midrule
		$\times$                                                       & 
		$\checkmark$                                                        & 
		$30^{\circ}$                                                            
		                                & 0 
		Mbps                                               \\
		\midrule
		$\checkmark$                                                   & 
		$\checkmark$                                                        & 
		$30^{\circ}$                                                            
		                                & 1683 
		Mbps        \\
		\bottomrule                                   
	\end{tabular}
\end{table}
Finally, the beam steering ability of the fabricated transmissive 
array is validated in this subsection. Compared to the communication settings 
in previous experiments, the transmit power is fixed to 13.6 dBm and the 
azimuth angle from the transmissive RIS to the transmitter is 
$\theta_t=$ 30°. The other settings are the same.
The experiment 
results are shown in Table VI.  Without the assistance of transmissive RIS, the 
data rate is 
reduced from 1024 Mbps to 450 Mbps due to the misalignment between the Tx and 
Rx horns. By introducing the transmissive RIS and scanning the beam it 
forms to 
the 30° direction, the maximum 1683 Mbps data rate is achieved. As a result, we 
can conclude that the transmissive RIS is able to dynamically steer its beam to 
adapt to the variation of the incident angle of the signal. Similar to the 
experiment in the previous subsection, the obstacle is again introduced into 
the communication prototype. The experiment results validate that the 
transmissive RIS is able to improve the data rate and overcome obstacles at an 
oblique incident angle $\theta_t = $ 30°. 

\section{Conclusion}
The transmissive RIS has a promising prospect in 6G communications to provide 
enhanced signal coverage in various communication scenarios. Compared to the 
reflective RIS, the research and experimental demonstration of the transmissive 
RIS and the transmissive RIS-aided communication system are considerably fewer. 
This work proposes a novel 2-bit transmissive RIS design with a 90° digital 
phase shifter and a 1-bit vertical current reversible dipole. The 2-bit phase 
controlling capability is achieved and measured. Experimental results show the 
transmissive RIS prototype with $16\times16$ elements can provide a broadside 
gain of 
22.0 dBi at 27.0 GHz. Moreover, a transmissive RIS-aided communication 
prototype is set up, and the system performances for three representative 
scenarios are measured. It is experimentally demonstrated that the transmissive 
RIS is able to achieve higher data rate, reduce the transmit power, improve the 
transmission capability through obstacles, and dynamically adapt to the signal 
propagation direction.
\bibliographystyle{IEEEtran}
\bibliography{IEEEabrv,refs}

\begin{thebibliography}{10}
\providecommand{\url}[1]{#1}
\csname url@samestyle\endcsname
\providecommand{\newblock}{\relax}
\providecommand{\bibinfo}[2]{#2}
\providecommand{\BIBentrySTDinterwordspacing}{\spaceskip=0pt\relax}
\providecommand{\BIBentryALTinterwordstretchfactor}{4}
\providecommand{\BIBentryALTinterwordspacing}{\spaceskip=\fontdimen2\font plus
\BIBentryALTinterwordstretchfactor\fontdimen3\font minus
  \fontdimen4\font\relax}
\providecommand{\BIBforeignlanguage}[2]{{%
\expandafter\ifx\csname l@#1\endcsname\relax
\typeout{** WARNING: IEEEtran.bst: No hyphenation pattern has been}%
\typeout{** loaded for the language `#1'. Using the pattern for}%
\typeout{** the default language instead.}%
\else
\language=\csname l@#1\endcsname
\fi
#2}}
\providecommand{\BIBdecl}{\relax}
\BIBdecl

\bibitem{1}
J.~Tang, S.~Xu, and F.~Yang, ``Design of a {2.5-D} 2-bit reconfigurable
  transmitarray element for {5G mmWave} applications,'' in \emph{Proc. 2020
  IEEE Int. Symp. Antennas and Propag.}, 2020, pp. 631--632.

\bibitem{2}
F.~E. Idachaba, ``{5G} networks: Open network architecture and densification
  strategies for beyond 1000x network capacity increase,'' in \emph{proc.
  Future Technol. Conf.}, 2016, pp. 1265--1269.

\bibitem{3}
T.~S. {Rappaport}, Y.~{Xing}, O.~{Kanhere}, S.~{Ju}, A.~{Madanayake},
  S.~{Mandal}, A.~{Alkhateeb}, and G.~C. {Trichopoulos}, ``Wireless
  communications and applications above 100 {GHz}: Opportunities and challenges
  for 6{G} and beyond,'' \emph{IEEE Access}, vol.~7, pp. 78\,729--78\,757, Jun.
  2019.

\bibitem{4}
W.~{Tang}, M.~Z. {Chen}, X.~{Chen}, J.~Y. {Dai}, Y.~{Han}, M.~{Di Renzo},
  Y.~{Zeng}, S.~{Jin}, Q.~{Cheng}, and T.~J. {Cui}, ``Wireless communications
  with reconfigurable intelligent surface: Path loss modeling and experimental
  measurement,'' \emph{IEEE Trans. Wireless Commun.}, vol.~20, no.~1, pp.
  421--439, Jan. 2021.

\bibitem{5}
P.~Wang, J.~Fang, X.~Yuan, Z.~Chen, and H.~Li, ``Intelligent reflecting
  surface-assisted millimeter wave communications: Joint active and passive
  precoding design,'' \emph{IEEE Trans. Veh. Tech.}, vol.~69, no.~12, pp.
  14\,960--14\,973, Dec. 2020.

\bibitem{6}
C.~Huang, A.~Zappone, G.~C. Alexandropoulos, M.~Debbah, and C.~Yuen,
  ``Reconfigurable intelligent surfaces for energy efficiency in wireless
  communication,'' \emph{IEEE Trans. Wireless Commun.}, vol.~18, no.~8, pp.
  4157--4170, Aug. 2019.

\bibitem{7}
L.~Dai, B.~Wang, M.~Wang, X.~Yang, J.~Tan, S.~Bi, S.~Xu, F.~Yang, Z.~Chen,
  M.~D. Renzo, C.-B. Chae, and L.~Hanzo, ``Reconfigurable intelligent
  surface-based wireless communications: Antenna design, prototyping, and
  experimental results,'' \emph{IEEE Access}, vol.~8, pp. 45\,913--45\,923,
  2020.

\bibitem{8}
Z.~Zhang and L.~Dai, ``A joint precoding framework for wideband reconfigurable
  intelligent surface-aided cell-free network,'' \emph{IEEE Trans. Signal
  Process.}, vol.~69, pp. 4085--4101, Jun. 2021.

\bibitem{9}
J.~Xu, Y.~Liu, X.~Mu, and O.~A. Dobre, ``{STAR-RISs}: Simultaneous transmitting
  and reflecting reconfigurable intelligent surfaces,'' \emph{IEEE Commun.
  Lett.}, vol.~25, no.~9, pp. 3134--3138, Sep. 2021.

\bibitem{10}
\BIBentryALTinterwordspacing
``{DOCOMO} conducts world’s first successful trial of transparent dynamic
  metasurface,'' 2020. [Online]. Available:
  \url{www.nttdocomo.co.jp/english/info/media center/pr/2020/0117 00.html}
\BIBentrySTDinterwordspacing

\bibitem{11}
M.~Sazegar, Y.~Zheng, C.~Kohler, H.~Maune, M.~Nikfalazar, J.~R. Binder, and
  R.~Jakoby, ``Beam steering transmitarray using tunable frequency selective
  surface with integrated ferroelectric varactors,'' \emph{IEEE Trans. Antennas
  Propag.}, vol.~60, no.~12, Dec. 2012.

\bibitem{12}
L.~Boccia, I.~Russo, G.~Amendola, and G.~Di~Massa, ``Multilayer antenna-filter
  antenna for beam-steering transmit-array applications,'' \emph{IEEE Trans.
  Microw. Theory Techn.}, vol.~60, no.~7, pp. 2287--2300, Jul. 2012.

\bibitem{13}
J.~Y. Lau and S.~V. Hum, ``Analysis and characterization of a multipole
  reconfigurable transmitarray element,'' \emph{IEEE Trans. Antennas Propag.},
  vol.~59, no.~1, pp. 70--79, 2011.

\bibitem{15}
W.~Pan, C.~Huang, X.~Ma, and X.~Luo, ``An amplifying tunable transmitarray
  element,'' \emph{IEEE Antennas Wireless Propag. Lett.}, vol.~13, pp.
  702--705, 2014.

\bibitem{14}
J.~Y. Lau and S.~V. Hum, ``Reconfigurable transmitarray design approaches for
  beamforming applications,'' \emph{IEEE Trans. Antennas Propag.}, vol.~60,
  no.~12, pp. 5679--5689, Dec. 2012.

\bibitem{16}
C.~Huang, W.~Pan, X.~Ma, B.~Zhao, J.~Cui, and X.~Luo, ``Using reconfigurable
  transmitarray to achieve beam-steering and polarization manipulation
  applications,'' \emph{IEEE Trans. Antennas Propag.}, vol.~63, no.~11, pp.
  4801--4810, Nov. 2015.

\bibitem{17}
M.~Frank, F.~Lurz, R.~Weigel, and A.~Koelpin, ``Electronically reconfigurable 6
  × 6 element transmitarray at {K-Band} based on unit cells with continuous
  phase range,'' \emph{IEEE Antennas Wireless Propag. Lett.}, vol.~18, pp.
  796--800, 2019.

\bibitem{18}
A.~Clemente, L.~Dussopt, R.~Sauleau, P.~Potier, and P.~Pouliguen, ``1-bit
  reconfigurable unit cell based on {PIN} diodes for transmit-array
  applications in {X}-band,'' \emph{IEEE Trans. Antennas Propag.}, vol.~60,
  no.~5, pp. 2260--2269, May 2012.

\bibitem{20}
W.~Pan, C.~Huang, X.~Ma, B.~Jiang, and X.~Luo, ``A dual linearly polarized
  transmitarray element with {1-Bit} phase resolution in {X}-band,'' \emph{IEEE
  Antennas Wireless Propag. Lett.}, vol.~14, pp. 167--170, 2015.

\bibitem{21}
L.~Di~Palma, A.~Clemente, L.~Dussopt, R.~Sauleau, P.~Potier, and P.~Pouliguen,
  ``Circularly-polarized reconfigurable transmitarray in {Ka}-band with beam
  scanning and polarization switching capabilities,'' \emph{IEEE Trans.
  Antennas Propag.}, vol.~65, no.~2, pp. 529--540, Feb. 2017.

\bibitem{23}
B.~D. Nguyen and C.~Pichot, ``Unit-cell loaded with {PIN} diodes for {1-Bit}
  linearly polarized reconfigurable transmitarrays,'' \emph{IEEE Antennas
  Wireless Propag. Lett.}, vol.~18, pp. 98--102, 2019.

\bibitem{24}
M.~Wang, S.~Xu, F.~Yang, and M.~Li, ``A {1-Bit} bidirectional reconfigurable
  transmit-reflect-array using a single-layer slot element with {PIN} diodes,''
  \emph{IEEE Trans. Antennas Propag.}, vol.~67, no.~9, pp. 6205--6210, Sep.
  2019.

\bibitem{22}
L.~Di~Palma, A.~Clemente, L.~Dussopt, R.~Sauleau, P.~Potier, and P.~Pouliguen,
  ``Experimental characterization of a circularly polarized 1 {Bit} unit cell
  for beam steerable transmitarrays at {Ka}-band,'' \emph{IEEE Trans. Antennas
  Propag.}, vol.~67, no.~2, pp. 1300--1305, Feb. 2019.

\bibitem{25}
Y.~Wang, S.~Xu, F.~Yang, and D.~H. Werner, ``1 {Bit} dual-linear polarized
  reconfigurable transmitarray antenna using asymmetric dipole elements with
  parasitic bypass dipoles,'' \emph{IEEE Trans. Antennas Propag.}, vol.~69,
  no.~2, pp. 1188--1192, Feb. 2021.

\bibitem{26}
C.-W. Luo, G.~Zhao, Y.-C. Jiao, G.-T. Chen, and Y.-D. Yan, ``Wideband 1 bit
  reconfigurable transmitarray antenna based on polarization rotation
  element,'' \emph{IEEE Antennas Wireless Propag. Lett.}, vol.~20, pp.
  798--802, 2021.

\bibitem{19}
A.~Clemente, L.~Dussopt, R.~Sauleau, P.~Potier, and P.~Pouliguen, ``Wideband
  400-element electronically reconfigurable transmitarray in {X} band,''
  \emph{IEEE Trans. Antennas Propag.}, vol.~61, no.~10, pp. 5017--5027, Oct.
  2013.

\bibitem{27}
C.-C. Cheng, B.~Lakshminarayanan, and A.~Abbaspour-Tamijani, ``A programmable
  lens-array antenna with monolithically integrated mems switches,'' \emph{IEEE
  Trans. Microw. Theory Techn.}, vol.~57, no.~8, Aug. 2009.

\bibitem{28}
F.~Diaby, A.~Clemente, R.~Sauleau, K.~T. Pham, and L.~Dussopt, ``2 bit
  reconfigurable unit-cell and electronically steerable transmitarray at {Ka}
  -band,'' \emph{IEEE Trans. Antennas Propag.}, vol.~68, no.~6, pp. 5003--5008,
  2020.

\bibitem{29}
B.~Wu, A.~Sutinjo, M.~E. Potter, and M.~Okoniewski, ``On the selection of the
  number of bits to control a dynamic digital {MEMS} reflectarray,'' \emph{IEEE
  Antennas Wireless Propag. Lett.}, vol.~7, pp. 183--186, 2008.

\bibitem{30}
H.~Yang, F.~Yang, S.~Xu, M.~Li, X.~Cao, J.~Gao, and Y.~Zheng, ``A study of
  phase quantization effects for reconfigurable reflectarray antennas,''
  \emph{IEEE Antennas Wireless Propag. Lett.}, vol.~16, pp. 302--305, 2017.

\bibitem{31}
Q.~Wu and R.~Zhang, ``Beamforming optimization for wireless network aided by
  intelligent reflecting surface with discrete phase shifts,'' \emph{IEEE
  Trans. Commun.}, vol.~68, no.~3, pp. 1838--1851, Mar. 2020.

\bibitem{32}
Y.~Xiao, F.~Yang, S.~Xu, M.~Li, K.~Zhu, and H.~Sun, ``Design and implementation
  of a wideband 1-bit transmitarray based on a {Yagi–Vivaldi} unit cell,''
  \emph{IEEE Trans. Antennas Propag.}, vol.~69, no.~7, pp. 4229--4234, Jul.
  2021.

\bibitem{33}
T.~Lambard, O.~Lafond, M.~Himdi, H.~Jeuland, and S.~Bolioli, ``A novel analog
  360° phase shifter design in {Ku and Ka} bands,'' in \emph{Proc. 4th Euro.
  Conf. Antennas Propag. (EuCAP)}, 2010, pp. 1--4.

\bibitem{35}
S.~X. Ta, H.~Choo, and I.~Park, ``Broadband printed-dipole antenna and its
  arrays for {5G} applications,'' \emph{IEEE Antennas Wireless Propag. Lett.},
  vol.~16, pp. 2183--2186, 2017.

\end{thebibliography}
\end{document}